# Problems and Consequences of Bilateral Notions of (Meta-)Derivability

Sara Ayhan[1]

## Abstract

A bilateralist take on proof-theoretic semantics can be understood as demanding of a proof system to display not only rules giving the connectives' provability conditions but also their refutability conditions. On such a view, then, a system with two derivability relations is obtained, which can be quite naturally expressed in a proof system of natural deduction but which faces obstacles in a sequent calculus representation. Since in a sequent calculus there are two derivability relations inherent, one expressed by the sequent sign and one by the horizontal lines holding *between sequents*, in a truly bilateral calculus both need to be dualized. While dualizing the sequent sign is rather straightforwardly corresponding to dualizing the horizontal lines in natural deduction, dualizing the horizontal lines in sequent calculus, uncovers problems that, as will be argued in this paper, shed light on deeper conceptual issues concerning an imbalance between the notions of proof vs. refutation. The roots of this problem will be further analyzed and possible solutions on how to retain a bilaterally desired balance in our system are presented.

## 1 Introduction

The philosophical background for the considerations in this paper is the view taken by what is nowadays called 'proof-theoretic semantics', more specifically by a *bilateralist* perspective on proof-theoretic semantics[1]. Proof-theoretic semantics (PTS)

---

[1] The term was coined by Schroeder-Heister in 1991, see also (Schroeder-Heister, 2024) for a detailed overview. The ideas, though, go back to and belong to the program of *general proof theory*, endorsed by, e.g., Prawitz (1971, 1973); Kreisel (1971); Martin-Löf (1975).

✉ Sara Ayhan
 sara.ayhan@rub.de

1 Fakultät für Philosophie und Erziehungswissenschaft, Ruhr University Bochum, Bochum, Germany

is situated in the broader program of *inferentialism* (see, e.g., Brandom, 2000), an account taking the meaning of linguistic expressions to be determined by how they are used in inferences. PTS is specifically concerned with the meanings of logical connectives and takes these, accordingly, to be determined by their rules given in a proof system, i.e., in that PTS is an account opposed to what is usually considered 'semantics' in logic, namely model-theoretic semantics. Traditionally, the proof systems considered in PTS are natural deduction systems (in Gentzen-style) but there are also approaches arguing for the possibility, the superiority, or even necessity of using sequent calculi for such an account (see, e.g., Dicher & Paoli, 2012a; Ripley, 2012b; Schroeder-Heister, 2000, 2015; Wansing, 2021).[2] Proof-theoretic bilateralism,[3] then, is the view that when it comes to meaning and consequence not only something like assertion, truth or provability conditions should be considered but equally the respectively dual concepts, i.e., denial, falsity or refutability conditions, which are taken as not reducible to the former but primitive, too. Thus, from a bilateralist viewpoint our usual proof systems, displaying only the former concepts, give an incomplete picture when we want them to determine the meaning of connectives, which means we need to extend them to display the dual concepts as well.

What I want to consider in this paper is how this would look like in a sequent calculus system.[4] Sequent calculi are interesting devices because they represent derivability on two levels, within sequents and between sequents. If one assumes that for a bilateral proof system we need dualized derivability in the form of proofs and refutations, as I will argue (Sect. 2), then it seems only reasonable to demand that in a sequent calculus *both* derivability relations must be dualized. Sketching out how such a sequent calculus could look like will unfold certain problems, though, namely an apparent imbalance between meta-provability and meta-refutability (Sect. 3). Therefore, I will analyze the reasons for this asymmetry and point out two possible solutions (Sect. 4), which will require us to revisit certain questions and notions that often seem to be taken for granted, like our conception of derivability, of constructiveness and assumptions about consistency. Thus, although I begin with a specific system of sequent calculus here to illustrate my points, the questions coming to light in the course of the paper are rather not of a technical but much more of a general and conceptual nature.

---

[2] The reasons why natural deduction (and specifically in Gentzen-style) has developed as the prevailing one in PTS are in my view historical, philosophical and technical but I do not deem the arguments appealing to sequent calculi as unreasonable, either. I will leave this discussion here aside and just assume that there is nothing in principle speaking against the idea that rules of a sequent calculus can be meaning-giving for the logical connectives.

[3] The term was coined in Rumfitt (2000), predecessors for the idea are, e.g., Price (1983); Smiley (1996); Humberstone (2000).

[4] Speaking of bilateralism within a sequent calculus framework, readers may think in the direction of what is sometimes called 'normative bilateralism' or what Ripley (Ripley, 2017) categorized as 'coherence-based bilateralism', which can be found in Restall (2005, 2013); Ripley (2013). This is very different from the present approach, though, in that there the form of bilateralism is inherent in reading or interpreting the sequent calculus as representing positions of assertion and denial, while here I am concerned with having explicit bilateral derivability.

## 2 Bilateralist Proof Systems

There are different approaches on how to implement bilateralism in PTS and the one that I will consider here is to do so on the level of derivability, i.e., our proof system is bilateral in that it displays two derivability relations, one for provability and one for refutability.[5] My motivation for this comes from what I think is a conceptual consequence of assuming a PTS-based approach, namely to consider logics as consequence relations and not as sets of theorems, which is just reasonable given that we need to consider *rule*-based proof systems in PTS, not axiomatic ones. On such a view, then, it seems reasonable to implement bilateralism on this very fundamental level of consequence relations.[6] It is also for this reason that I will consider a specific logic here, namely the bi-intuitionistic logic 2Int (Wansing, 2016a), i.e., a conservative extension of intuitionistic logic by a connective dual to implication, called 'co-implication'. 2Int works with two consequence relations, which are displayed in its proof systems by comprising *proof rules* and *refutation rules* as well as both *assumptions* (taken to be verified) and *counterassumptions* (taken to be falsified). Note that there is also a use of 'bi-intuitionistic logic' in the literature to refer to a specific system, namely BiInt, also called 'Heyting-Brouwer logic' (see, e.g., Goré, 2000; Kowalski & Ono, 2017; Postniece, 2010; Rauszer, 1974). Except for having the same language, the systems crucially differ, though, in the interpretation of co-implication, in their validity concepts and in their constructiveness. BiInt does not work with two consequence relations and does not enjoy the disjunction property, saying that if $A \vee B$ is provable, either $A$ is provable or $B$ is provable, nor the dual conjunction property, saying that if $A \wedge B$ is refutable, either $A$ is refutable or $B$ is refutable (Pinto and Uustalu, 2018), while 2Int has both.[7] Thus, 2Int is a constructive system, which is also a desirable property from a PTS-standpoint. I will restrict details about 2Int to what is necessary for understanding the purpose of the present paper, since this is, as I said above, more about conceptual than technical questions.[8]

The proof systems of interest for the present matter are a natural deduction system, N2Int, introduced in Wansing (2017), and a corresponding cut-free sequent calculus, SC2Int (Ayhan, 2021). In N2Int the proof rules are indicated by using single lines, while the refutation rules are indicated by using double lines and as premises an ordered pair $(\Gamma; \Delta)$ (with $\Gamma$ and $\Delta$ being finite, possibly empty sets) of assumptions ($\Gamma$) and counterassumptions ($\Delta$) is used. It should be emphasized that given the constructive properties of 2Int both proving and refuting is to be understood in a constructive manner in this context. Since the term 'refutation' is often used in a much weaker way, meaning something like 'showing invalidity',[9] the most important point to stress here is probably the following: Understanding refutation constructively amounts to something along the lines of showing logical falsity, also called 'direct falsification'. The notion of proof can be thought of as in usual intuitionistic logic expressed by the Brouwer-Heyting-Kolmogorov (BHK) clauses. It is important to bear in mind, though, that in a bilateralist context, 'constructive' cannot be equated with 'intuitionistic'. The intuitionistic notion of refutation, as implying falsity, is much weaker than if we consider it also as a *primitive* notion. This is what is done in Nelson (1949) and López-Escobar (1972), in the latter of which counterparts of the BHK clauses for refutation are expressed. Let us therefore briefly reconsider both the BHK clauses and their dual versions for refutation, which are here slightly adapted from the version to be found in (López-Escobar, 1972, p. 363). The BHK clauses give an interpretation of complex formulas in terms of proof via the clauses i)-iv) below. For atomic formulas it is assumed that we have an intuitive understanding of what constitutes a proof.

---

[5] For an overview and examples of the different approaches to bilateralism, see respectively the introduction and the contributions in Ayhan (2023) as well as Ripley (2020).

[6] This is not to say that other approaches for bilateralist (readings of) proof systems are not reasonable. For the purpose of this paper I will leave these aside, though.

[7] It is for this reason that I would not call BiInt a constructive system. In this context an anonymous reviewer raised the question about what makes a system constructive and whether the disjunction property is enough. Although I think this is a very interesting question, I do not intend to delve into a deeper discussion on this here because it extends the scope of this paper. In my opinion, though, the disjunction property is a necessary (which is why BiInt cannot be called constructive) but not sufficient condition for a system to be constructive for the following reason: in such a bilateralist context as is the case for 2Int, I think it makes sense to also require the dual conjunction property to hold. Likewise, if we have a first-order system, then it seems to be reasonable to also demand the existence property to hold, etc.

[8] It has been remarked that using 2Int or in general a bi-intuitionistic logic, here is odd because co-implication does not play any further role for the paper. This is true, however, it is for the two features that I sketched briefly above, namely implementing two bilateral consequence relations in the proof system and being constructive, that I deem 2Int the best choice for what I want to show.

i) A proof of $A \wedge B$ is a pair $\langle d, e \rangle$, where $d$ is a proof of $A$ and $e$ is a proof of $B$.
ii) A proof of $A \vee B$ is a pair $\langle i, d \rangle$, where $i = 0$ and $d$ is a proof of $A$, or $i = 1$ and $d$ is a proof of $B$.
iii) A proof of $A \rightarrow B$ is a function that maps any proof of $A$ to a proof of $B$.
iv) There is no proof of $\bot$.

For the corresponding dual versions, the notion of refutation is taken to be primitive and complex formulas are to be understood in terms of this, expressed by the clauses i)$^d$-iv)$^d$ below. As in the case of the BHK clauses, for atomic formulas an intuitive understanding of what constitutes a refutation is presupposed.

i)$^d$ A refutation of $A \wedge B$ is a pair $\langle i, d \rangle$, where $i = 0$ and $d$ is a refutation of $A$, or $i = 1$ and $d$ is a refutation of $B$.
ii)$^d$ A refutation of $A \vee B$ is a pair $\langle d, e \rangle$, where $d$ is a refutation of $A$ and $e$ is a refutation of $B$.
iii)$^d$ A refutation of $A \rightarrow B$ is a pair $\langle d, e \rangle$, where $d$ is a proof of $A$ and $e$ is a refutation of $B$.
iv)$^d$ There is no refutation of $\top$.

To see how a natural deduction system with rules both for proving and for refuting works, we can consider the following rules for conjunction as an example.[10]

---

[9] See my comments and the references in fn 16 below.

[10] Note that derivations can also consist of a mixture of applications of proof and refutation rules, which becomes apparent when the rules for (co-)implication are considered. For the sake of conciseness, I will

Introduction rules for $\wedge$ in N2Int:

$$
\begin{array}{cccc}
(\Gamma;\Delta) & (\Gamma';\Delta') & (\Gamma;\Delta) & (\Gamma;\Delta) \\
\vdots & \vdots & \vdots & \vdots \\
\underline{A \quad\quad B} \\
A \wedge B
\end{array} \wedge I
\quad
\dfrac{\overline{\overline{A}}}{A \wedge B} \wedge I_1^d
\quad
\dfrac{\overline{\overline{B}}}{A \wedge B} \wedge I_2^d
$$

As an anonymous reviewer remarked, it might be worthwhile to point out how these rules relate to what is probably one of the most famous bilateralist systems, namely Rumfitt's (Rumfitt, 2000) natural deduction calculus with signed formulas standing for speech acts of assertion and denial. Structurally it is very similar in that $\overline{A}$ in N2Int can be read as $+A$ ('asserting $A$') in Rumfitt's system and $\overline{\overline{A}}$ in N2Int as $-A$ ('denying $A$') in Rumfitt's system. However, from the philosophical interpretation I do think those systems are rather different, one being concerned with speech acts and one with derivability relations. And of course, when considering the full system of rules, the big difference is that Rumfitt presents a calculus for classical logic, while N2Int one for a bi-intuitionistic logic.

To turn to the corresponding sequent calculus now, in SC2Int the bilateralism is implemented by having a signed sequent sign $\vdash^*$ (with $* \in \{+, -\}$) expressing *derivability* or *dual derivability* from a *context* $(\Gamma; \Delta)$ on the left side of the sequent (with $\Gamma$ and $\Delta$ defined as before). If $\Gamma = \Delta = \emptyset$, then we can also speak of *provability* and *refutability*, i.e., these are the stronger notions. Let's consider again part of the rules for conjunction to see the comparison to the natural deduction calculus.

Right introduction rules for $\wedge$ in SC2Int:[11]

$$\dfrac{(\Gamma;\Delta) \vdash^+ A \quad (\Gamma;\Delta) \vdash^+ B}{(\Gamma;\Delta) \vdash^+ A \wedge B} \wedge R_1^+$$

$$\dfrac{(\Gamma;\Delta) \vdash^- A}{(\Gamma;\Delta) \vdash^- A \wedge B} \wedge R_{1a}^- \quad\quad \dfrac{(\Gamma;\Delta) \vdash^- B}{(\Gamma;\Delta) \vdash^- A \wedge B} \wedge R_{1b}^-$$

A usual reading of natural deduction (ND) systems when compared to sequent calculi (SC) is that, roughly speaking, the horizontal lines in ND translate to the sequent signs in SC. This seems in accordance to what happens in this bilateralist setting, where the single lines in N2Int correspond to $\vdash^+$ in SC2Int and the double lines to $\vdash^-$. However, what about the horizontal lines in SC2Int? After all, in sequent calculi we have two sorts of derivability relations expressed, one between (sets of)

---

only consider conjunction here because this connective has the 'simplest' rules but already these are sufficient to show in the next section the problems that would equally occur with the other connectives. For the same reason I will restrict my examples in the next section to only the right introduction rules in sequent calculus: Just for these we will need to consider 12 rules and the problems becoming apparent in this setting would equally occur for the left introduction rules.

[11] In SC2Int these are the only right introduction rules for conjunction; the subscript $_1$ is only added to make reference easier in the next section where I add more rules of this kind. For the full system including the structural rules, see (Ayhan, 2021). The structural rules are not important for the purpose of the present paper, though, which is why I leave them out here.

formulas within sequents, which is expressed by the sequent sign, and one between sequents, expressed by the horizontal lines. It is evident that these remain as usual, though, i.e., there is no duality of single and double lines here. So – and that might be the worry from a bilateralist point of view – don't we keep a unilateral element in a calculus like `SC2Int`? Don't we, in order to be 'truly' bilateral in our sequent calculus, also have to consider a duality of derivability relations on the vertical level, and thus, introduce single *and* double lines in the sequent calculus, too, next to having $\vdash^+$ and $\vdash^-$?

## 3 Sketching a Truly Bilateral Sequent Calculus

So, how could such a 'fully' bilateral sequent calculus look like? To get started, all I want to do in this section is give a sketch of what one *might* expect such a calculus to look like, i.e., it will be more intuition pumping at this point than deep argumentation. Thus, I have to ask the readers to bear with me for the time being and trust that certain questions that may be arising here will become clearer in the later sections.

If we go along with the idea of having two different horizontal lines in SC, what we need next to rules expressing that a sequent is *proven*, ending like this

$$\frac{}{(\Gamma; \Delta) \vdash^+ A \wedge B} \wedge R^+ \qquad \frac{}{(\Gamma; \Delta) \vdash^- A \wedge B} \wedge R^-$$

are also ones expressing that a sequent is *refuted*, ending like this

$$\frac{}{\overline{(\Gamma; \Delta) \vdash^+ A \wedge B}} \wedge R^+ \qquad \frac{}{\overline{(\Gamma; \Delta) \vdash^- A \wedge B}} \wedge R^-.$$

Henceforth, for convenience, I will call sequents of the former kind, i.e., with a single line on top, simply 'single-lined' and sequents of the latter kind 'double-lined'. For the single-lined sequents, I have given the premises above. Now, for the double-lined version of $\wedge R^+$ it seems adequate to consider the same premises as the ones from $\wedge R_{1a}^-$ and $\wedge R_{1b}^-$ because in these the premises express a guarantee to get a proof of the sequent $(\Gamma; \Delta) \vdash^- A \wedge B$, so then they should equally guarantee a refutation of the sequent $(\Gamma; \Delta) \vdash^+ A \wedge B$:

$$\frac{\overline{(\Gamma; \Delta) \vdash^- A}}{\overline{(\Gamma; \Delta) \vdash^+ A \wedge B}} \wedge R_{2a}^+ \qquad \frac{\overline{(\Gamma; \Delta) \vdash^- B}}{\overline{(\Gamma; \Delta) \vdash^+ A \wedge B}} \wedge R_{2b}^+.$$

Reading the horizontal lines in the same way as they are to be read in ND as expressing proof and refutation, $\wedge R_{2a}^+$ would be read as saying that from a proof of a sequent of the form $(\Gamma; \Delta) \vdash^- A$, i.e., expressing that there is a refutation of $A$ from $(\Gamma; \Delta)$, we get a refutation of the sequent $(\Gamma; \Delta) \vdash^+ A \wedge B$, i.e., expressing that there is a proof of $A \wedge B$ from $(\Gamma; \Delta)$. This makes sense because for a proof of $A \wedge B$ we would need a proof of both $A$ and $B$. So, if we have proven that the context $(\Gamma; \Delta)$ leads to a refutation of one of these formulas, then surely this seems to refute the possibility to get a proof of $A \wedge B$ from this same context.

Now, it seems that if the conclusion can be introduced by double lines, the same should be possible for the premises, so we should also consider the following rules, which are the result of dualizing the premises of $\wedge R_{2a}^+$ and $\wedge R_{2b}^+$, switching the single-lined $\vdash^-$ into double-lined $\vdash^+$:

$$\frac{\overline{(\Gamma;\Delta) \vdash^+ A}}{(\Gamma;\Delta) \vdash^+ A \wedge B} \wedge R_{3a}^+ \qquad \frac{\overline{(\Gamma;\Delta) \vdash^+ B}}{(\Gamma;\Delta) \vdash^+ A \wedge B} \wedge R_{3b}^+$$

These seem reasonable as well, expressing that if it has been refuted that the sequent $(\Gamma;\Delta) \vdash^+ A$, resp. $(\Gamma;\Delta) \vdash^+ B$, holds, this refutes the sequent $(\Gamma;\Delta) \vdash^+ A \wedge B$, too.

So, the next question would be if we shouldn't get the same kind of dualizing for the other direction by switching a single-lined $\vdash^+$ to a double-lined $\vdash^-$ in our $\wedge R_1^+$ rule. However, here is where we get into trouble. Consider the following rule, which would be the outcome of this:[12]

$$\frac{\overline{(\Gamma;\Delta) \vdash^- A} \quad \overline{(\Gamma;\Delta) \vdash^- B}}{(\Gamma;\Delta) \vdash^+ A \wedge B} \wedge R_{\sharp}^+$$

Accepting this rule would mean to say that refuting that $(\Gamma;\Delta)$ leads to refuting $A$ and refuting that $(\Gamma;\Delta)$ leads to refuting $B$ would be sufficient to *prove* that $(\Gamma;\Delta)$ gives a proof of $A \wedge B$. But that does not seem to be sufficient at all, at least, and this is what I consider here to be the case, if our notion of proof is in any way a *constructive* one. Just refuting that there is a refutation of something cannot give us a constructive *proof* of something. So, it seems that from the 'structural looks' of it, we should have a rule like $\wedge R_{\sharp}^+$, because – on the bilateralist assumption that these notions are on a par – why should the dualizing of the implementations of proof and refutation only work in one direction? When thinking about what the interpretation of this rule would amount to, though, it seems that we must dismiss it. What is problematic from a bilateralist point of view, then, is that this leads to an asymmetry because we *do* have single *and* double-lined premises for the double-lined conclusion of $(\Gamma;\Delta) \vdash^+ A \wedge B$ but only a rule with single-lined premises is possible for the same single-lined conclusion. So, it seems possible to go from proofs to proofs *or* refutations, but from refutations only to refutations, *not* to proofs.

Maybe, however, this is due to the fact that we are introducing a sequent with $\vdash^+$, expressing provability, so let's go through this same procedure for introducing $A \wedge B$ on the right of $\vdash^-$ and see whether this time, for a sequent expressing refutability, we get the same kind of asymmetry but the other way around – a restriction for going from single-lined premises to double-lined conclusions. So, again, we already have one set of rules, which we can take from SC2Int, with single-lined conclusions and single-lined premises and the question is how premises for a double-lined conclusion could look like. Based on our procedure above, it seems to make sense again to take the same premises as from rule $\wedge R_1^+$, since it would mean to commit to saying that having proven both proofs of $A$ and of $B$ from $(\Gamma;\Delta)$, this refutes the possibility of a refutation of $A \wedge B$ from $(\Gamma;\Delta)$:

---

[12] The subscript $\sharp$ will be used to indicate all rules that I consider as at least *prima facie* problematic.

$$\dfrac{\overline{(\Gamma;\Delta) \vdash^+ A} \quad \overline{(\Gamma;\Delta) \vdash^+ B}}{(\Gamma;\Delta) \vdash^- A \wedge B} \wedge R_3^-$$

For a refutation of $A \wedge B$ we would need at least one refutation, of $A$ or of $B$, so surely, if we have *proven* both formulas from the same context, this means that we cannot get a refutation of them and so, also not of $A \wedge B$.

By dualizing these premises as we did before, changing single-lined $\vdash^+$ into double-lined $\vdash^-$, we get another rule again:

$$\dfrac{\overline{\overline{(\Gamma;\Delta) \vdash^- A}} \quad \overline{\overline{(\Gamma;\Delta) \vdash^- B}}}{(\Gamma;\Delta) \vdash^- A \wedge B} \wedge R_4^-$$

This rule seems unproblematic as well. After all, if we have refuted both the possibility of refuting $A$ and of refuting $B$ from context $(\Gamma;\Delta)$, then this refutes getting a refutation of $A \wedge B$ from this context for the same reason as before: we would need at least the refutation of one of these formulas in order to get a refutation of their conjunction. However, considering rules $\wedge R_{1a}^-$ and $\wedge R_{1b}^-$ now, having single-lined conclusions and trying the same dualization of the premises here, we run into problems again:

$$\dfrac{\overline{\overline{(\Gamma;\Delta) \vdash^+ A}}}{(\Gamma;\Delta) \vdash^- A \wedge B} \wedge R_?^- \qquad \dfrac{\overline{\overline{(\Gamma;\Delta) \vdash^+ B}}}{(\Gamma;\Delta) \vdash^- A \wedge B} \wedge R_?^-$$

Having these rules in our system seems problematic because it would mean committing to having refuted that there is a proof of $A$ (or $B$) from a context as sufficient for proving that there is a refutation of $A \wedge B$ from this context. Intuitively – given that we are in a constructive setting – these premises do not seem to guarantee such a conclusion, though. What we need for a constructive refutation of $A \wedge B$ is a refutation of $A$ or of $B$; having refuted that there is a proof of one of those formulas simply does not amount to that, since there could be neither.

So, we get the same kind of asymmetry here again: While with rules starting from single-lined premises it seems acceptable to infer single-lined *and* double-lined conclusions, with rules starting from double-lined premises only double-lined conclusions can be inferred. Rules going from double-lined to single-lined sequents seem unacceptable. Thus, it appears that refutation is somehow a weaker concept than proof, since less can be deduced from having a refutation than can be from having a proof. Or, to put it differently, proof is the stronger concept because more restrictions apply for guaranteeing a proof than for guaranteeing a refutation. So, what we wanted to achieve with this double-lined sequent calculus was *more bilaterality* in our system by considering the two derivability relations of proving and refuting not only within sequents but also between sequents. What we ended up with, though, is *less bilaterality* because proof and refutation do not seem to be on a par in this setting.

## 4 Reasons for the Asymmetry Between Proofs and Refutations and Possible Solutions

The problem, or even 'dilemma' is that our aim was to make our sequent calculus more bilateral but somehow this exposed proof as a stronger notion than refutation in this setting and thus we got less bilaterality. What I want to show in this final section, is that although this may seem like a rather technical problem and also one that only comes up in this specific context of bilateralist PTS, it actually sheds light on deeper philosophical and conceptual issues. I will just reveal, though, what may be the roots of the problem and on this basis point out possible directions as ways out of the dilemma. I will not settle on a definite answer or show how we can get a 'fully' bilateral sequent calculus after all, since this would require much more technical work exceeding the scope of this paper.

### 4.1 Making the Background Assumptions Explicit

What I have done so far was in part merely appealing to intuitions about what 'seems reasonable' as rules and what not. So let us make two background assumptions explicit, which I think are prevalent here and which abstract away from any specific rules for logical connectives. It should be emphasized that these intuitions and background assumptions are not what I ultimately deem as correct. Rather, in Sect. 4.3 I will explain where and why the thoughts outlined here go wrong in my opinion. Thus, some readers may not agree with these intuitions but in general (in conversations or Q & A's at conferences, etc.) I found that they did seem intuitive to many people, which is why I start out from this point of view. The first one is a dependency assumption that I think is quite common for most logicians, saying that it cannot be the case that both $(\Gamma; \Delta) \vdash^+ A$ and $(\Gamma; \Delta) \vdash^- A$ hold.[13] I call it a dependency assumption because – and this has been claimed several times when I presented this material – many people seem to have a strong intuition that the notions of proof and refutation must be somehow connected in the sense that they must exclude each other. So, there seems to be a certain common ground that these notions are not completely independent: you cannot claim that there is a proof but that this does not mean that there could not *also* be a refutation of the same expression. And the other way around: claiming that there exists a refutation for an expression means for many scholars that this precludes there being a proof of that same expression. I want to note that in an earlier version of this paper I called this a 'consistency assumption', not completely without reason: For lots of people these two issues would be strongly interconnected, also for López-Escobar (López-Escobar, 1972), as we will see below (Sect. 4.3.1). This dependency in the sense of mutual exclusivity means some kind of consistency insurance: there cannot be both a proof and a refutation of the same expression. So, if we say that it cannot be the case that both $(\Gamma; \Delta) \vdash^+ A$ and $(\Gamma; \Delta) \vdash^- A$, this means that if one of these sequents is proven to hold, this precludes the other to hold, i.e., having a proof of one is sufficient for having refuted the other. Formalizing this condition would amount to the following structural rules:[14]

---

[13] As an anonymous reviewer correctly pointed out, in 2Int this is actually not a justified assumption unless the left side of the sequents is empty, i.e., while it is not possible to both prove and refute the same formula, it is possible to both derive and dually derive the same formula from the same context, namely in case the context is inconsistent. This would be the case if the same formula is part of the assumptions and the counterassumptions, which is not excluded in 2Int. Since most people are not too familiar with the specifications of this system, though, I wanted to capture here rather the intuitions which I found logicians more commonly share.

$$\frac{(\Gamma; \Delta) \vdash^+ A}{(\Gamma; \Delta) \vdash^- A} S_{1a} \qquad \frac{(\Gamma; \Delta) \vdash^- A}{(\Gamma; \Delta) \vdash^+ A} S_{1b}$$

Thus, a proof of a derivation (resp. of a dual derivation) of $A$ from $(\Gamma; \Delta)$ is sufficient for a refutation of a dual derivation (resp. of a derivation) of $A$ from $(\Gamma; \Delta)$.

The second background assumption at work here is that, on the other hand, it *can* be the case that neither $(\Gamma; \Delta) \vdash^+ A$ nor $(\Gamma; \Delta) \vdash^- A$ hold. This is also a very common assumption especially in constructive settings because otherwise it would mean to commit to decidability for every formula. It seems intuitive, though, to assume that there may be formulas which are neither provable nor refutable in our system. This means that having refuted that one of these sequents holds does not amount to having a proof that the other must hold, i.e., formalized, the following does *not* hold:

$$\frac{\overline{(\Gamma; \Delta) \vdash^+ A}}{(\Gamma; \Delta) \vdash^- A} S_{2a\sharp} \qquad \frac{\overline{(\Gamma; \Delta) \vdash^- A}}{(\Gamma; \Delta) \vdash^+ A} S_{2b\sharp}$$

Thus, a refutation of a derivation (resp. of a dual derivation) of $A$ from $(\Gamma; \Delta)$ is *not* sufficient for a proof of a dual derivation (resp. of a derivation) of $A$ from $(\Gamma; \Delta)$. Hence, this is where we have a mismatch, an apparent asymmetry between proofs and refutations.

### 4.2 Possible Solution: Reconsidering Our Conception of Sequent Calculus

One possible cause of the asymmetry and thus, approach to this problem is our conception of sequent calculi in general. Depending on that, we could get a possible solution, namely by taking a strict stance on SC merely being a meta-theory of ND calculi.[15] This would suggest to conceive of ND systems as the *actual* way of capturing derivability and SC simply as a description of that. Thus, a rule like $\wedge R^-_{1a}$ from SC2Int would be understood as expressing in a meta-language "there is a derivation in ND, such that [$\wedge I^d_1$ from N2Int]", not more. On such an understanding it could be argued, then, that it is unnecessary to have more than one horizontal line in SC for the simple reason that the horizontal lines in a bilateral ND system for proving and refuting already translate into $\vdash^+$ and $\vdash^-$, so, there is not more to translate.[16] So, one way out of this dilemma would be to insist that a calculus like SC2Int is already perfectly bilateral as it is and that there is no need to make it more bilateral than that. An objection to this, however, is if one argues that by now SC systems have developed into proof systems independent of ND, meaning that they should be considered as calculi in their own right. It seems, then, that the horizontal lines in SC represent derivability just like in ND, although it is between different linguistic expressions. This would simply reflect that derivability need not only exist between (sets of) formulas but can also exist between, e.g., arguments, i.e., entities already containing derivations.[17]

---

[14] One can also think of these as *coordination principles* between proofs and refutations, something that is common to specify in bilateralist systems, see, e.g., Rumfitt (2000).

[15] This is not an unusual stance at all, see, e.g., (Prawitz, 1965, p. 90) (Hacking, 1979, p. 292) or (Dummett, 1991, pp. 40, 185f.).

### 4.3 Possible Solution: Reconsidering Our Notion of Constructiveness for Proofs and Refutations

Therefore, what I think is rather the root of the problem is that in our reading of SC rules as given in Sect. 3 our notions of proof and refutation are understood as constructive only on the level within sequents, i.e., for what it means to prove or refute formulas, but not on the level between sequents. I will speak of *meta-provability* and *meta-refutability* henceforth when I speak of these latter derivability relations indicated by the horizontal lines in a sequent calculus.[18] When I said above that the rules $S_{1a}$ and $S_{1b}$ are fine, while $S_{2a\natural}$ and $S_{2b\natural}$ are problematic, it seems that our conception of meta-provability and meta-refutability was not understood equally constructive within these rules. To be precise, our conception of meta-refutation seems non-constructive in that case, while our conception of meta-proof is constructive.[19] I will depict in more detail what I mean by this below but it should be noted that this is related to another issue that I want to clarify. In such a calculus with rules displaying both single and double horizontal lines, not being able to prove a sequent does not equate with necessarily being able to refute this sequent. Just as in a bilateralist natural deduction system we may be able to neither prove nor refute a formula – because neither our proof nor our refutation rules give us enough means to do so – the same can be the case for such a sequent system: We do not have a bivalence in the sense that the only question to ask is "Is that sequent provable or not?" but additionally we can ask the question "Is that sequent refutable or not?". From a positive answer to either of these questions, it may be the case that we can draw further conclusions (or not, we will see these two possibilities below) but it seems rather clear that no conclusions can be drawn from a negative answer to either of these questions.

---

[16] In a presentation of this material it has been remarked that, even on such an understanding of SC, it might still make sense to have the two kinds of horizontal lines, where we would understand the double lines simply as saying "there is no derivation, such that...". This seems to go in the direction of what researchers like Goranko, Pulcini, Skura or Varzi have worked on, e.g., (Goranko, 2020, Goranko, Pulcini, & Skura, 2019, Varzi & Pulcini, in press) where the notion of refutation they are first and foremost interested in is that of deriving non-valid formulas (although in the latter they call these 'rejection' systems, while 'refutation' seems closer to our notion here). However, this is very different from the present notion of refutation, which is that of showing definite falsity.

[17] The use in natural language would support this: it does not seem odd to say something like "from this argument follows...". See also (Dicher and Paoli, 2021, pp. 635f.) for more objections against the metatheoretical view.

[18] In that, the present paper relates to work that has recently been done on the notion of *meta-inferentiality* see, e.g., (Pailos & Da Ré 2023) but they are not concerned with (meta-)refutation, while this is the more crucial point here. Also in accordance with this is a very early paper in the bilateralist PTS tradition (of course, none of these terms were in use at that time) by von Kutschera ([1969]2025), who attempts to give generalized rule schemata for proofs and refutations and for this considers not only sequents as expressing provability and refutability but, trying to make it as general as possible, also expressions in the form of nested sequents displaying these relations *between* sequents.

[19] This might relate to a point an anonymous reviewer raised, namely that the 'problematic' rules look similar to what could be expressed by a sort of meta-linguistic double negation elimination kind of reasoning in that "a negative valence operator applied to a carrier of negative semantic value is transformed into a positive valence structure". If we reject double negation elimination because of an implicit intuitionistic 'positive' bias, then this might be another explanation why one would deem those rules as prima facie more counterintuitive than the ones leading from positive to negative values.

So, first of all, at least from a bilateralist point of view it is demanded that we conceive meta-refutability just as constructive as meta-provability. We still have two different ways to proceed from here in order to restore the symmetry. One option is to assume that basically proofs and refutations consist of the *same underlying construction*, in which case we should also accept rules $S_{2a\frac{1}{2}}$ and $S_{2b\frac{1}{2}}$. The other option is to assume that proofs and refutations have *independent* underlying constructions, in which case neither $S_{2a\frac{1}{2}}$ and $S_{2b\frac{1}{2}}$ *nor* $S_{1a}$ and $S_{1b}$ seem justified anymore. Note that in both cases my claim is that the first background assumption, i.e., the dependency assumption, that I sketched out in Sect. 4.1 is at heart of the issue here. The difference is that if we go with the first option, we would have to commit to a different understanding or interpretation of this dependency than what we formulated before, while if we go with the second option, we have to outright dismiss this dependency assumption. This was only a summary of the options, which I hope will become clearer with the explanations in the next sections.

### 4.3.1 Proofs and Refutations as Unified Constructions

Let us consider the first option. Under a constructive notion of (meta-) refutability, this can be considered plausible with a reasoning along the lines that can already be found in López-Escobar (1972), for example, where constructive interpretations of negative statements in the form of refutations are considered:

> Suppose that $c$ is a construction that refutes $\dashv A$. That is $c$ refutes that $A$ is refutable. Hence from $c$ we must be able to extract the information that no construction will ever refute $A$. Since we are assuming that no construction can both refute and prove the same formula and we know from $c$ that no construction will ever refute $A$, it appears reasonable to stipulate that the construction $c$ then proves $A$. (It could also be argued that the only way in which the construction $c$ could encode the information that there will never be found a refutation of $A$ is to encode a proof of $A$). (López-Escobar, 1972, p. 364)

Although López-Escobar is concerned with what it means to refute a negative statement in this paragraph, since he understands negation itself as a refutation,[20] this comes down to the question what it means to refute a proof, resp. a refutation, i.e., it is what I am concerned with here when asking what it means to refute a sequent that expresses a proof, resp. a refutation from the left to the right side. Transferring these claims to our context would mean that it is reasonable to assume that refuting a proof is equivalent to proving a refutation, while refuting a refutation is equivalent to proving a proof. Thus, considering proofs and refutations both as kinds of inferences, here we seem to be speaking of meta-inferences. Given that the connective of implication is often understood as expressing a notion of inference in the object language, it may be worthwhile to consider our conception of implication here in order to make our conception of inferences in general clearer. This can seem like reversing the reasoning because our definition of implication should be derived from our understanding of inferences, but I think that actually these conceptions are just intertwined in such a way that it does not make sense to be too strict about the concrete starting point.

---

[20] "A construction $c$ proves $\dashv A$ iff $c$ refutes $A$" (López-Escobar, 1972, p. 364). Note that he makes the same two background assumptions as we did in Sect. 4.1: "Of course it is possible for a given construction

Let us say that we subscribe to a constructive understanding of implication in the sense that is given by these BHK/López-Escobar clauses from above:

iii)     A proof of $A \to B$ is a function that maps any proof of $A$ to a proof of $B$.
iii)$^d$    A refutation of $A \to B$ is a pair $\langle d, e \rangle$, where $d$ is a proof of $A$ and $e$ is a refutation of $B$.[21]

Then it seems reasonable to transfer this onto our understanding of inferences in general, i.e., to understand

$$\overline{(\Gamma; \Delta) \vdash^+ A}$$

as refuting a positive inference from the left to the right side of the sequent, which would come down to presenting a proof of the left and a refutation of the right side. This seems to be exactly the same information, though, that we would need for proving a negative inference from the left to the right side of the sequent, i.e., for deriving

$$\overline{(\Gamma; \Delta) \vdash^- A}$$

Similarly, if we think about what construction would justify deriving

$$\overline{(\Gamma; \Delta) \vdash^+ A},$$

---

to neither prove nor refute a formula. On the other hand we do not think that it is reasonable to allow that a given construction $c$ both proves and refutes a formula $A$".

[21] An anonymous reviewer remarked that this looks distinctly non-constructive because it amounts to the material conditional. However, I think the non-constructivity of classical logic is rather due to an interplay with non-constructive notions of other connectives (specifically of disjunction with respect to proof and of conjunction with respect to refutation). After all, this notion of refuting implication is considered by López-Escobar (López-Escobar, 1972), explicitly interested in constructivity, and by Nelson for his constructive logic N4. Also, note that other conceptions of refuting an implication are possible, such as a connexive one: A refutation of $A \to B$ is a construction that converts a proof of $A$ into a refutation of $B$. It is not important here to decide on one or the other but it should be mentioned that *if* one would decide on the connexive understanding, it would only be consequential to implement the same interpretation for the notion of refutation, both within sequents and between sequents. Thus, the symmetry would also hold in this case.

this seems to be a construction in the sense of iii), i.e., converting any proof of $(\Gamma; \Delta)$ into a proof of $A$.[22] And this also seems to be exactly the kind of information that we would need to refute a negative inference from $(\Gamma; \Delta)$ to $A$, i.e., for deriving

$$\overline{(\Gamma; \Delta) \vdash^- A}.$$

After all, if we can retrieve this kind of information from a construction, this seems to refute that a refutation in the sense of iii)$^d$, containing a proof of $(\Gamma; \Delta)$ and a refutation of $A$, can be presented.

Thus, given these equivalences, the following rules (formerly indexed with $\frac{1}{4}$) would indeed be fine:

$$\frac{\overline{(\Gamma; \Delta) \vdash^+ A}}{(\Gamma; \Delta) \vdash^- A} S_{2a} \qquad \frac{\overline{(\Gamma; \Delta) \vdash^- A}}{(\Gamma; \Delta) \vdash^+ A} S_{2b}$$

The symmetry is restored, then, because just like proving a proof (refutation) of $A$ is sufficient for refuting a refutation (proof) of $A$, as is stated in $S_{1a}$ and $S_{1b}$, in the same way refuting a proof (refutation) of $A$ is sufficient for proving a refutation (proof) of $A$. To recapitulate what was the wrong understanding of the dependency assumption from Sect. 4.1: It is not so much about one construction (proof or refutation) of an expression cancelling out the possibility of a dual construction for the same expression. Rather, it is about assuming a dependency between proofs and refutations in that essentially the same information can be encoded in different ways; having one way is sufficient for having the other then, which is why it is allowed to flip back and forth between proofs and refutations of dual sequents. Note that the second background assumption still holds, i.e., it can be the case that we have neither a proof nor a refutation of $(\Gamma; \Delta) \vdash^* A$ in our system. It is just settled that *if* we have one or the other, then this means that we have a construction encoding the same information that we need to prove or refute the dual sequent.

### 4.3.2 Proofs and Refutations as Independent Constructions

Let us consider the second option now, i.e., conceiving of proofs and refutations as independent constructions. As said above, this means now not to assume a different understanding of the dependency assumption but to drop it altogether. This is also where an important difference to López-Escobar's remarks comes into play: He assumes that there cannot be a construction that both proves and refutes $A$. However, what he does not consider is the possibility that there might be a construction proving $A$, while there also being *another* construction refuting $A$. If we drop any dependency assumption between proofs and refutations, though, it seems unjustified not to consider the possibility of such a case. Importantly, both understandings of the dependency assumption are dropped here. The one from Sect. 4.1, meaning that having a proof (respectively, a refutation) for an expression does not preclude the existence of there also being a refutation (respectively, a proof) for the same expression. There can just be another, completely independent construction; why not? And secondly, also the reading of dependency from Sect. 4.3.1 is dropped: Assuming a complete independence between proofs and refutations, it seems unjustified to assume a joint underlying construction that can transform proofs into refutations and vice versa.

---

[22] I did not make explicit what it means to be a proof of a pair of assumptions and counterassumptions but intuitively it would mean to prove everything in $\Gamma$ and refute everything in $\Delta$.

What we need to be careful about here is distinguishing provability vs. refutability and derivability vs. dual derivability because this has consequences for how this relates to the consistency of the calculus. If we assume this independence of constructions for proofs and refutations of formulas, that is, what we would be concerned with in a natural deduction system or what would be expressed by derivations of empty left-sided sequents in a sequent calculus, then this means at least in principle to dispense with consistency, too, because it could be the case then that we get both a proof and a refutation of $A$. Surely this possibility will be counter-intuitive and worrisome for many people. It should be noted, though, that this would only be reason to worry if our system is trivializing in the face of such inconsistencies and that there are so-called *contradictory logics*, which do exhibit exactly this feature: they contain formulas of the form $A$ and $\sim A$ in their set of theorems but are not trivial systems in the sense that every formula is provable.[23] On the other hand, if it is about (dual) derivability, i.e., with open premises, this could be expressed as having both $(\Gamma; \Delta) \vdash^+ A$ and $(\Gamma; \Delta) \vdash^- A$. This, however, is a much weaker form of inconsistency and it is actually something unproblematic in the system 2Int because it can just occur if the same formulas appear both in the assumptions and the counterassumptions. In model-theoretic terms, as Wansing (Wansing, 2016a, p. 442) describes it, in 2Int support of truth and support of falsity are not mutually exclusive and thus, it can be the case that a certain state supports both the truth and the falsity of the same formula.

What does this mean for the structural rules considered above? Even though our usual intuitions may tell us that we cannot get both $(\Gamma; \Delta) \vdash^+ A$ and $(\Gamma; \Delta) \vdash^- A$, without a dependency assumption, this is not even *not* ruled out but an actual possibility in a calculus like 2Int.[24] So, proving $(\Gamma; \Delta) \vdash^+ A$ neither precludes $(\Gamma; \Delta) \vdash^- A$ (countering the first reading of the dependency assumption, as given in Sect. 4.1) nor does it give us a transformation into a constructive refutation of its dual sequent (countering the second reading of the dependency assumption, as given in Sect. 4.3.1), and vice versa for the opposite polarities. The consequence of this would be of course that we need to reconsider rules $S_{1a}$ and $S_{1b}$ now and conclude that these are just as problematic as we thought of the rules $S_{2a\natural}$ and $S_{2b\natural}$ in Section 4.1. So, basically all structural rules expressing some kind of coordination principles between proofs and refutations have to be considered unjustified under such a conception. Thus, as opposed to the previous section where it was argued to accept all of the structural rules under a certain interpretation, here the desired symmetry between proofs and refutations is restored by *rejecting* all of the rules:

$$\dfrac{\overline{(\Gamma;\Delta) \vdash^+ A}}{(\Gamma;\Delta) \vdash^- A} S_{1a\not{\,}} \qquad \dfrac{\overline{(\Gamma;\Delta) \vdash^- A}}{(\Gamma;\Delta) \vdash^+ A} S_{1b\not{\,}} \qquad \dfrac{\overline{(\Gamma;\Delta) \vdash^+ A}}{(\Gamma;\Delta) \vdash^- A} S_{2a\not{\,}} \qquad \dfrac{\overline{(\Gamma;\Delta) \vdash^- A}}{(\Gamma;\Delta) \vdash^+ A} S_{2b\not{\,}}.$$

---

[23] See, e.g., (Niki, 2024; Wansing, in press). Although most of these systems use this way of expressing inconsistency, the provability of $\sim A$ can be easily translated to refutability of $A$ and thus to the form of bilateralism which I prefer in this paper. Note that this does not mean to drop a constructive understanding of proofs and refutations, as is made explicit in Wansing (2024).

[24] As is most famously and constantly argued by Priest (see, e.g., Priest (1998, 2006) or Priest et al. (2023) for further references), we can simply take this as a hint to question our 'intuitive' background assumptions, face the fact that from centuries of indoctrination in the spirit of Aristotelian and then classical logic our intuitions are very much biased toward not accepting inconsistencies and maybe conclude that this is not a firm ground to argue against these.

Given that we assume proofs and refutations to be in principle completely independent constructions, it just seems reasonable not to have any general coordination principles between them.

## 5 Conclusion and Outlook

To conclude, there are good reasons to consider proof systems with two derivability relations in the realm of bilateralism. The question that arises for a sequent calculus is then: should these relations be displayed only on the level of sequents or also between sequents? The obstacle that appears when choosing the 'more bilateral' option by implementing the duality of derivability relations on both levels, though, is that with seemingly more bilaterality on one end we get less bilaterality on the other end, to wit, an asymmetry between proofs and refutations in the form that more seems derivable from the former than from the latter. What these considerations shed light on are conceptual and to date rather undeveloped issues on how to understand the derivability relations in sequent calculi and thus, meta-concepts of proof and refutation.[25] When we consider sequent calculi to be proof systems in their own right, i.e., not just as meta-conceptions of natural deduction systems, then there are two paths to retain symmetrical *and* constructive notions of proof and refutation also on a meta-level. Either we keep an assumption of dependency between proofs and refutations, which results in assuming a joint underlying construction, i.e., a close connection between these notions. Then a formulation of symmetrical coordination principles seems justified. Or we admit the possibility of separate and truly independent constructions underlying proofs and refutations, which may lead to give up certain assumptions concerning consistency. In this case, though, any coordination principles between proofs and refutations seem ill-advised, which also restores the symmetry between these notions.

As it was not my aim to give a full account of how a 'truly' bilateral sequent calculus could look like after all these considerations, I will leave this issue here with just a remark toward a possible solution. There is a logic, 2C Wansing (2016b), which is very close to 2Int in its vocabulary and in its bilateral representation but which is a *bi-connexive* logic, meaning that the interpretation of implication and co-implication differs crucially and this has an important effect: 2C is a contradictory logic in the sense that it is not only paraconsistent but its set of theorems actually contains formulas of the form $A$ and $\sim A$. As such, this might be a suitable system to go with the second option outlined in Sect. 4.3.2. I will leave it for future work to determine whether a truly bilateralist representation of this logic in the sense as it was outlined in this paper is possible.

---

[25] A remarkable exception being the paper by Dicher & Paoli (Dicher and Paoli, 2021) who also mention some further rare exceptions.

**Acknowledgements** I would like to thank Hitoshi Omori for bringing up the question of `truly' bilateral sequent calculi during a presentation of mine on related material and for the following extensive discussions we had on that topic. Thanks go also to two anonymous reviewers for their thorough and constructive comments.

**Funding** Open Access funding enabled and organized by Projekt DEAL. This research has received funding from the European Research Council (ERC) under the European Union's Horizon 2020 research and innovation programme, grant agreement ERC-2020-ADG, 101018280, ConLog.